\documentstyle[12pt,fleqn,cite,epsfig]{article}

\begin{document}
\topmargin 0pt
\oddsidemargin 0mm

\begin{titlepage}
\begin{flushright}
IP/BBSR/2002-25\\
hep-th/0210230
\end{flushright}

\vspace{5mm}
\begin{center}
{\Large \bf D-Branes at angle in pp-wave Background}
\vspace{6mm}

{\large
Rashmi R. Nayak}\\  
\vspace{5mm}
{\em Institute of Physics\\ 
Bhubaneswar 751 005\\ India\\
\vspace{3mm}

email: rashmi@iopb.res.in}

\end{center}
\vspace{5mm}
\centerline{{\bf{Abstract}}}
\vspace{5mm}
We show the existence of classical solutions of a system
of $D3$-branes oriented at an arbitrary angle with respect to each other, 
in a six dimensional pp-wave background obtained from
$AdS_3\times S^3\times R^4$, with $NS-NS$ and $R-R$ $3$-form 
field strength. These $D$-brane bound states are shown to 
preserve $1/16$ of the supersymmetries.
We also present more $D$-brane bound state solutions by applying 
T-duality symmetries. Finally, the probe analysis is discussed 
along with a brief outline of the open string construction.

\end{titlepage}

\newpage

Recently, the study of string theory in PP-wave background
has been a subject of intense discussion. It is known 
that ``Penrose'' limit 
plays an important role, in obtaining such solutions,
as any Einstein gravity admits a plane 
wave background through these limits\cite{penrose}. The  
supergravity realization of string theory in pp-wave background
through these limits are studied as well\cite{tsyt,guven,blau,
blau1,blau2,billo}.String theory in this background is easy to 
handle due to the presence 
of a natural light cone gauge and are exactly solvable
in Green-Schwarz formalism\cite{metsaev2,tsyt2,russo}. The exact 
solvability of string theory in these backgrounds, provide a powerful
tool to investigate the AdS/CFT correspondence in a better way
\cite{malda,mukhi,sonn}. The maximally supersymmetric type IIB
pp-wave background was found in \cite{blau}, on which string theory is 
exactly solvable and has been used to show the duality between string 
and gauge theories, by means of an one to one mapping of 
the stringy modes to the operaters in the gauge theory.

D-branes, known as non-perturbative and extended objects, 
also survive in this limit. The study of $D$-branes 
in pp-wave background, arising from a
geometry of $AdS_p \times S^q$ type,  has been a subject of wide 
interest\cite{blau2}.
Explicit supergravity solutions of these objects and 
their open string spectrum, have been discussed
at length in the
literature\cite{dabh,lee,kumar,sken,singh,bain,alishah,
pal,michi,alok,stelle}. These objects play an 
important role in understanding the duality between string and gauge 
theories. BPS (supersymmetric) bound of these objects 
have also been useful in understanding the physics of black holes  
in string theory. It is now known that, there are also 
supersymmetric bound states where the component branes
are at relative angles, with the angles being restricted to lie in  
an $SU(N)$ subgroup of rotations\cite{douglas}. Explicit classical 
solution, to the supergravity equations coresponding to 
the configurations of branes at angle,  
are already discussed in the literature\cite{gauntlett,cvetic,
myers,hambli,leigh}. These bound states are of wide applications,
in understanding various dualities of 
superstring theories as well as the study of 
black holes in string theories.  
 
In veiw of the importance of these bound states, in this paper 
we show the existence of the classical solution of a system
of $D$-branes oriented at angle, belongs to $SU(2)$ subgroup of 
rotations, in pp-wave background
with constant $NS-NS$ $3$-form field strengths.
In particular, we present two 
supergravity solutions, one coresponds to a single $D3$-brane and the
other one is a system of $D3$-branes which are oriented 
by certain $SU(2)$ angle, as described above, in $NS-NS$ plane wave
backgrounds. Some other $D$-brane bound 
states are also obtained by applying $T$-duality and 
$S$-duality symmetries in the above supergravity solution.
It has been explicitly shown that,
these bound states preserve $1/16$ of the supersymmetries. These brane 
configurations are also analyzed from the point of view of probe branes. 
Finally, we have discussed the open string construction briefly.   

We now begin by writing down the classical solution of a $D3$-brane rotated 
by certain $SU(2)$ angle in the pp-wave background of $AdS_3\times S^3
\times R^4$\cite{russo,michi,alok} with constant $NS-NS$ $3$-form field 
strength:
\begin{eqnarray}
ds^2 &=&\sqrt{1+X_1}\Bigg\{{1 \over 1+X_1}\bigg(
2{d x^+}{d x^-} - {\mu^2}\sum_{i=1}^4 {x^2_i~ {(d x^+)^2}} \cr
& \cr
&+& [1 + X_1 \cos^2\alpha][(d x^5)^2 + (d x^7)^2] 
+ [1 + X_1 \sin^2\alpha][(d x^6)^2 + (d x^8)^2] \cr
& \cr
&+& 2X_1 \sin\alpha \cos\alpha (d x^7 d x^8 -d x^5 d x^6) \bigg)
+ \sum_{i=1}^4 (d x^i)^2\Bigg\} \cr
& \cr
H_{+12} &=& H_{+34} = 2\mu , \cr
& \cr
F^{(5)}_{+-68i} &=& - {\partial_i X_1\over (1 + X_1)^2}\cos^2\alpha,~~~~  
F^{(5)}_{+-67i} =  {\partial_i X_1\over (1 + X_1)^2}
\cos\alpha \sin\alpha, \cr
& \cr
F^{(5)}_{+-57i} &=& {\partial_i X_1\over (1 + X_1)^2}\sin^2\alpha,~~~~
F^{(5)}_{+-58i} = - {\partial_i X_1\over (1 + X_1)^2}
\cos\alpha \sin\alpha, \cr  
& \cr
e^{2 \phi_b} &=& 1.
\label{d3-1}
\end{eqnarray}
and $X_1$ is given by
\begin{equation}
X_1(\vec r) = { 1\over 2}
\bigg( {\ell_1 \over \vert \vec r - \vec r_1\vert}\bigg)^2.\label{d3-1a}
\end{equation}
Where r is the radius vector in the transverse space, 
defined as ${r^2 = \sum_{i=1}^4 (x^i)^2}$, $r_1$ is the location 
of $D3$-brane in transvers space and  $X_1$ is the Harmonic function,
which satisfies the Green function equation in the transverse space.\\   

To start with, the $D3$-brane is along $x^+, x^-, x^6$ and
$x^8$. By applying a rotation 
between $x^5-x^6$ and $x^7-x^8$ planes
following \cite{myers},  with rotation angles $(\alpha_1, \alpha_2) =
(0, \alpha)$, we get the configuration where the original 
$D3$-brane is tilted by an angle $\alpha$. 
In the limit $\mu = 0$, the above solution reduces to the
flat space $D3$-brane solution rotated by $SU(2)$ angle, 
which can be obtained  
by applying a $T$-duality transformation, along one of the 
transverse coordinate of the rotated $D2$-membrane 
solution as given in\cite{hambli}. One can also check 
that in the limit $\alpha = 0$, the above solution exactly matches 
with the $D3$-brane solution given in\cite{alok}.
Type IIB field equations (see e.g.\cite{duff}) are being satisfied by the 
solution given in eqn.(\ref{d3-1}). For example, the constant $3$-form 
field strengths along the pp-wave directions are in fact needed to 
cancel the $\mu$-dependent part of $R_{++}$ equation. One can also 
find out  the rotated $D3$-brane solution in constant $R-R$ three
form pp-wave background by applying $S$-duality symmetry in the above
solution. Under $S$-duality transformation the above solution will
remain as it is with the constant $NS-NS$  $3$-form field strength
being replaced to $R-R$ $3$-form field strength.   

Now we present the low energy classical solution of a system 
of two $D3$ branes, when both are placed at a relative $SU(2)$ 
angle with respect to each other. The supergravity solution of 
such a system in pp-wave background with constant $NS-NS$ $3$-form
field strength is given by:  

\begin{eqnarray}
ds^2 &=& \sqrt{1+X}\Bigg\{{1 \over 1+X}\bigg(
2{d x^+}{d x^-} - {\mu^2}\sum_{i=1}^4 {x^2_i~{(d x^+)^2}} \cr
& \cr 
&+& (1+X_2)\big[(d x^5)^2 + (d x^7)^2\big] + (d  x^6)^2 + (d x^8)^2 \cr
& \cr
&+& X_1\left[( \cos \alpha d x^5 - \sin \alpha d x^6)^2
+ ( \cos \alpha d x^7 + \sin \alpha d x^8)^2\right]\bigg)
+ \sum_{i=1}^4 (d x^i)^2\Bigg\},\cr
& \cr
H_{+12} &=& H_{+34} = 2 \mu, \cr
& \cr
F^{(5)}_{+-68i} &=& {\partial_i\Big\{{{X_2 + X_1 \cos^2\alpha +
X_1 X_2 \sin^2\alpha} \over {(1 + X)}}\Big\}},\cr
& \cr
F^{(5)}_{+-58i} &=& - F^{(5)}_{+-67i} = {\partial_i\Big\{{{X_1 \cos\alpha 
\sin\alpha} \over {(1 + X)}}\Big\}},\cr
& \cr
F^{(5)}_{+-57i} &=& - {\partial_i\Big\{{{(X_1+ X_1 X_2) \sin^2\alpha}
 \over {(1 + X)}}\Big\}},~~~~e^{2 \phi_b} = 1 .
\label{d3-2}
\end{eqnarray}
and $X$ is given by
\begin{equation}
X =\, X_1 + X_2 + X_1 X_2 \sin^2 \alpha,\label{d3-2a}
\end{equation}
where as defined earlier, $X_{1,2} = { 1\over 2}
\bigg( {\ell_{1,2} \over \vert \vec r - \vec r_{1,2}\vert}\bigg)^2$

In this case, to start with the two $D3$-branes are parallel to each
other and are lying along $x^+, x^-, x^6$, $x^8$ . By applying 
an $SU(2)$ rotation, as described in previous case,
the second brane rotated by an angle $\alpha$, now
lies in $x^+, x^-, x^5$ and $x^7$. At the same time the branes are   
delocalized along $x^5-x^7$ and $x^6-x^8$ planes respectively.
One can check that the solution given in eqn.(\ref{d3-2}), solves type IIB
field equations. In the limit $\mu = 0$, the above solution goes back  
to that given in \cite{myers}. By setting $X_2 = 0$, the above
solution reduces to the one given in eqn.(\ref{d3-1}). We would like to 
point out that at $\alpha = \pi/2$, the solution given in eqn.(\ref{d3-2})
represents the supergravity solution of two $D3$-branes 
intersecting orthogonally along a string. We 
emphasise that the possibility of this configuration is already
discussed in\cite{alok},  after applying successive T-dualities along 
the $D1-D5$ solution\cite{alok}. 
Therefore the solutions given in eqn.(\ref{d3-1})   
and eqn.(\ref{d3-2}), gives a pp-wave generalization of the ones given
in\cite{myers,hambli}. 

We now apply $T$-duality transformation along the world volume 
directions of the configuration given in eqn. (\ref{d3-2}) 
to generate more interesting  $D$-brane bound states in pp-wave
background. For example, applying $T$-duality along $x^8$, one gets:
   
\begin{eqnarray}
ds^2 &=&\sqrt{1 + X}\Bigg\{{1 \over 1 + X}\bigg(
2{d x^+}{d x^-} - {\mu^2}\sum_{i=1}^4 {x^2_i~ {(d x^+)^2}} \cr
& \cr
&+& (1 + X_2 )(d x^5)^2 + (d x^6)^2  
+ X_1(\cos\alpha d x^5 - \sin\alpha d x^6)^2 \bigg) \cr
& \cr
&+& {(d x^7)^2 + (d x^8)^2 \over 1 + X_1 \sin^2\alpha}  
+ \sum_{i=1}^4 (d x^i)^2\Bigg\} \cr
& \cr
H_{+12} &=& H_{+34} = 2\mu , \cr
& \cr
F^{(4)}_{+-6i} &=& {\partial_i\Big\{{{X_2 + X_1 \cos^2\alpha +
X_1 X_2 \sin^2\alpha} \over {(1 + X)}}\Big\}},~~~~  
F^{(4)}_{+-5i} =  {\partial_i\Big\{{{X_1 \cos\alpha 
\sin\alpha} \over {(1 + X)}}\Big\}},\cr
& \cr
B_{78} &=& - {X_1\cos\alpha \sin\alpha \over 1 + X_1 \sin^2\alpha},~~~~
e^{2\phi_a} = {\sqrt{1 + X}\over 1 + X_1 \sin^2\alpha}.\label{d2-d4}
\end{eqnarray}

One notices that setting $X_1 = 0$, one gets a $D2$-brane lying along the
coordinates,  $x^+$, $x^-$ and $x^6$ and at the same time 
`delocalized' along $x^5$, $x^7$ and $x^8$. Setting $X_2 = 0$, the 
solution given in eqn. (\ref{d2-d4}), reduces to the $D2-D4$ bound 
state in pp-wave background. By applying $S$-duality in the solution
given in eqn.(\ref{d3-2}), one can get classical solution
of a  system of $D3$-branes at an angle 
in $R-R$ pp-wave background. Again by applying $T$-duality 
transformation along the directions $x^6$ and $x^8$, one gets:
   
\begin{eqnarray}
ds^2 &=&\sqrt{1 + X}\Bigg\{{1 \over 1 + X}\bigg(
2{d x^+}{d x^-} - {\mu^2}\sum_{i=1}^4 {x^2_i~ {(d x^+)^2}}\bigg) \cr
& \cr
&+& {(d x^5)+ (d x^6) + (d x^7)^2 + (d x^8)^2 \over 1 + X_1 \sin^2\alpha}  
+ \sum_{i=1}^4 (d x^i)^2\Bigg\} \cr
& \cr
F_{+1268} &=& F_{+3468} = 2\mu , \cr
& \cr  
F^{(3)}_{+-i} &=& - {\partial_i\Big\{{{X_2 + X_1 \cos^2\alpha +
X_1 X_2 \sin^2\alpha} \over {(1 + X)}}\Big\}},~~~~  
F^{(5)}_{+-78i} = {\partial_i\Big\{{{X_1 \cos\alpha 
\sin\alpha} \over {(1 + X)}}\Big\}},\cr
& \cr
B_{56} &=& - B_{78} = {X_1\cos\alpha \sin\alpha \over 1 + X_1
\sin^2\alpha},~~~~
e^{2\phi_a} = {{1 + X}\over (1 + X_1 \sin^2\alpha)^2}.\label{dx-dy}
\end{eqnarray}
   
Now one can see that by setting $X_1 = 0$ the above solution 
reduces to that of a $D$-string lying along $x^+, x^-$ and 
delocalised along $x^5,..,x^8$ directions. Setting $X_2 = 0$  
the above supergravity solution coresponds to a $D(5,3,3,1)$
bound state in pp-wave background with constant $R-R$ $5$-form 
field strengths. One can see that at $\mu = 0$ limit 
this solution reduces to one given in \cite{myers}, which 
can be obtained by applying $T$-duality trasformation along 
one of the transverse directions. 
Similarly, by applying $S$-duality and $T$-duality
transformation in eqn.(\ref{d3-2}),
one can generate more bound states of various $D$-branes in 
$NS-NS$ and $R-R$ plane-wave backgrounds.
We however, skip the details here.  
We again emphasize that the bound state solutions presented 
here also gives the pp-wave generalization of
the ones given in flat space\cite{myers,myers1,costa}.

It has already been discussed in the literature\cite{douglas} that
a system of $D$-branes oriented at certain angle, lying inside the 
$SU(N)$ subgroup of rotations, with respect 
to each other, preserve certain amount of unbroken supersymmetries. 
We now confirm this fact, in pp-wave background through the examples 
discussed in this paper by analyzing the type IIB supersymmetry 
variations explicitly. The supersymmetry variation of dilatino 
and gravitino fields of type IIB supergravity in ten dimension,
in string frame, is given by \cite{john,fawad}:
\begin{eqnarray}
\delta \lambda_{\pm} &=& {1\over2}(\Gamma^{\mu}\partial_{\mu}\phi \mp
{1\over 12} \Gamma^{\mu \nu \rho}H_{\mu \nu \rho})\epsilon_{\pm} + {1\over
  2}e^{\phi}(\pm \Gamma^{M}F^{(1)}_{M} + {1\over 12} \Gamma^{\mu \nu
  \rho}F^{(3)}_{\mu \nu \rho})\epsilon_{\mp},
\label{dilatino}
\end{eqnarray}
\begin{eqnarray}
\delta {\Psi^{\pm}_{\mu}} &=& \Big[\partial_{\mu} + {1\over 4}(w_{\mu
  \hat a \hat b} \mp {1\over 2} H_{\mu \hat{a}
  \hat{b}})\Gamma^{\hat{a}\hat{b}}\Big]\epsilon_{\pm} \cr
& \cr
&+& {1\over 8}e^{\phi}\Big[\mp \Gamma^{\mu}F^{(1)}_{\mu} - {1\over 3!}
\Gamma^{\mu \nu \rho}F^{(3)}_{\mu \nu \rho} \mp {1\over 2.5!}
\Gamma^{\mu \nu \rho \alpha \beta}F^{(5)}_{\mu \nu \rho \alpha
  \beta}\Big]\Gamma_{\mu}\epsilon_{\mp},
\label{gravitino}
\end{eqnarray}

where we have used $(\mu, \nu, \rho)$ to describe the ten dimensinal 
space-time indices, and hat's represent the corresponding tangent space 
indices. Solving the above two equations for the solution describing
the system of two  $D3$-branes as given in eqn. (\ref{d3-2}), we get
several conditions on the spinors. First the dilatino variation gives:
 
\begin{equation}
\Gamma^{\hat +}(\Gamma^{\hat 1\hat 2} + \Gamma^{\hat 3\hat
  4})\epsilon_{\pm} = 0.
\end{equation}
Gravitino variations gives the following conditions on the spinors:
\begin{eqnarray}
\delta \psi_+^{\pm} &\equiv &\partial_{+}\epsilon_{\pm} -
{{\partial_i X\over8(1 +
X)^{3/2}}}\Gamma^{\hat-\hat+}\epsilon_{\pm} ~{\mp}~  {\mu\over2\sqrt{1 +
X}}(\Gamma^{\hat1\hat2} + \Gamma^{\hat3\hat4})\epsilon_{\pm} \cr
& \cr
&{\mp}& {1\over 8} ~\Gamma^{\hat +\hat -\hat 6\hat 8\hat i}~\Bigg[
{{(1 + X_1 \sin^2 \alpha)^2 ~{\partial_i X_2} + \cos^2\alpha ~{\partial_i X_1}}
\over {(1 + X)^{3/2}(1 + X_1 \sin^2 \alpha)}}\Bigg]
\Gamma^{\hat -} \epsilon_{\mp} \cr
& \cr
&{\mp}& {1\over 8}  \Gamma^{\hat +\hat -\hat 5\hat 7\hat i}~\Bigg[
{1\over (1 + X)^{5/2}}(1 + X_1 \sin^2 \alpha) \cr
& \cr
&\times& ({{X_1}^2 \cos^2\alpha \sin^2\alpha }~{\partial_i X_2} + 
(1 + X_2)^2 \sin^2 \alpha ~{\partial_i X_1}) \cr 
& \cr
&-& {{({X_1}^2 \cos^2\alpha \sin^2\alpha)(~(1 + X_1 \sin^2 \alpha)^2 
~{\partial_i X_2} + \cos^2\alpha ~{\partial_i X_1})}\over (1 + X)^{5/2}
(1 + X_1 \sin^2 \alpha)} \cr
& \cr 
&+& {1\over{(1 + X)^{5/2}}}
\bigg(2{X_1}^2 \cos^2\alpha \sin^2\alpha ~{\partial_i X_2} + 
2{X_1}^3 \cos^2\alpha \sin^4\alpha ~{\partial_i X_1} \cr
& \cr
&-& 2 X_1(1 + X_2)\cos^2\alpha \sin^2\alpha ~{\partial_i X_1}\bigg)
\Bigg]\Gamma^{\hat -} \epsilon_{\mp}  \cr
& \cr 
&{\mp}& {1\over 8}~\bigg\{{{\Gamma^{\hat +\hat -\hat 5\hat 8\hat i}~
-~\Gamma^{\hat +\hat -\hat 6\hat 7\hat i}}
\over {(1 + X)^2( 1 + X_1 \sin^2\alpha)}}\bigg\} 
\Bigg[(- X_1\cos\alpha \sin \alpha ~{\partial_i X_2} \cr
& \cr
&+& (1 + X_2)\sin\alpha
\cos\alpha ~{\partial_i X_1} - {X_1}^2 \sin^3\alpha \cos\alpha
 ~{\partial_i X_1})( 1 + X_1 \sin^2\alpha) \cr
& \cr 
&+& X_1 \cos\alpha \sin\alpha
( 1 + X_1 \sin^2\alpha)^2~{\partial_i X_2} \cr
& \cr 
&+& X_2\cos^3\alpha \sin\alpha ~{\partial_i X_1}\Bigg] 
\Gamma^{\hat -} \epsilon_{\mp} = 0, \label{susy}
\end{eqnarray}
\begin{equation}
\delta \psi_-^{\pm} \equiv \partial_{-}\epsilon_{\pm}=0. \label{susy2}
\end{equation}
In writing down the above variations we have made use of a necessary
condition: 
\begin{equation}
\Gamma^{\hat +} \epsilon_{\pm} = 0,\label{lightcone}
\end{equation}
Now, gravitino variation (\ref{susy}) is solved by imposing:
\begin{eqnarray}
(\Gamma^{\hat 5\hat 8} - \Gamma^{\hat 6\hat
7})\epsilon_{\mp} = 0,~~~~(\Gamma^{\hat 5\hat 7} + \Gamma^{\hat 6\hat
8})\epsilon_{\mp} = 0, \label{rotation}
\end{eqnarray}
\begin{eqnarray} 
\Gamma^{\hat +\hat -\hat 6\hat 8}\epsilon_{\mp} = \epsilon_{\pm},~~~~
\Gamma^{\hat +\hat -\hat 5\hat 7}\epsilon_{\mp} = \epsilon_{\pm},\label{brane}
\end{eqnarray}
in addition to the condition:
\begin{equation}
(1 - \Gamma^{\hat 1\hat 2\hat 3\hat 4})\epsilon_{\pm} = 0. \label{pp}
\end{equation}
To explain it further, below, we state the cancellations clearly.  
Imposing the condition (\ref{rotation}),
the last term of the gravitino variation (\ref{susy}) vanishes and
the fourth and fifth terms add up to give the
coefficent of the second term. Imposition of the brane 
conditions (\ref{brane}) along with (\ref{pp}), solves gravitino 
variation (\ref{susy}) fully.        
The expressions for the other two gravitino variations: 
$\delta \psi_a^{\pm}$  $(a = 5,...,8)$ and $\delta \psi_i^{\pm}$ 
$(i = 1,...,4)$, are of similar type and they are also 
solved by following the same arguments described above.
Therefore, the configuration of two $D3$-branes with a relative $SU(2)$
angle between the two, as given in eqn.(\ref{d3-2}), preserves $1/16$ 
supersymmetry. However the supersymetry obtained here is consistent
with the $D3$-brane supersymmetry given in \cite{alok}, when two of 
them are parallel to each other.
The above supersymmetry can also be found out by doing 
an analysis following \cite{douglas,townsend,ohta}, by choosing the
`rotation matrix' appropriately. For convenience, 
the expressions for the non vanishing Christoffels and some of the spin
connections are presented in Appendix A.

We now consider the situation from worldvolume 
point of view, where we will use a close relation between  
$\kappa$-symmetry and supersymmetry to determine the 
fraction of supersymmetry preserved by the above 
described configuration. We will use the $\kappa$-symmetry formulation of 
$Dp$-brane\cite{agan,wall,berg,ortin,chu} to show this. 
To do this computation, we will treat
the branes, involved in the intersection, as probes, propagating  
in $D=10$ space-time of the type 
$AdS_3 \times S^3 \times R^4$ with a six dimentional pp-wave. 
As described in \cite{ortin}, we have a system of 
two $D3$-branes intersecting at arbitrary angle with a vanishing $2$-form
BI field. For simplicity, each brane is identified with a subspace 
of a ten dimentional spacetime.  

Given a $D$-brane probe the surviving supersymmetries satisfy 
the condition:

\begin{equation}
(1 - \Gamma)\xi = 0,
\label{k1}
\end{equation}
\noindent
where $\Gamma$ is a projector that depends on the details of the
brane configuration. For vanishing BI field,  $\Gamma$ will take a 
form as defined in \cite{ortin}:

\begin{equation}
\Gamma = \Gamma^{\prime}_{(0)}, 
\end{equation}
\noindent
where $\Gamma^{\prime}_{(0)}$ and $\Gamma_{(0)}$ are as follows:   

\begin{equation} 
\Gamma^{\prime}_{(0)} = (\sigma_3)^{p-3\over 2}
{i \sigma_2 \otimes\Gamma_{(0)}},\>\>\> (for\>type\>IIB)
\end{equation}
with
\begin{equation}
\Gamma_{(0)} = {1 \over (p+1)!~{\sqrt -detG} }\epsilon^{i_1......i_p+1}
\partial_{i_1}X^{M_1}.......\partial_{i_{p+1}}X^{M_{p+1}}
\Gamma^{\prime}_{M_1}....._{M_{p+1}}.
\end{equation}
$\Gamma^{\prime}_M$ are the ten dimentional $\Gamma$-matrices in 
a coordinate basis defined by 

\begin{equation} 
\Gamma^{\prime}_M = {E^A}_M \Gamma_A,
\end{equation}
\noindent
with $\Gamma_A$ bieng the flat space matrices.
The induced world volume matric $G_{ij}$ is:
\begin{equation}
G_{ij} = {\partial_i} X^M {\partial_j} X^N g_{MN}.
\end{equation}
\noindent
For our case, the first $D3$-brane is lying along ${(x^+,x^-,x^6,x^8)}$
and stuck at the origin. Imposing the light cone gauge and physical gauge
conditions: $X^+(\tau,\xi) = p^+ \tau$, $X^-(\tau,\xi) = {\xi}^1$, and
$X^k(\tau,\xi) = {\xi}^k$, as explained in \cite{chu}, and denoting
the world volume coordinates of the $D$-brane by $(\tau,\xi^k)$,
$k= 1,6,8$, we have:    
\begin{equation}
\Gamma_{(0)} = \Gamma_{\hat + \hat - \hat 6 \hat 8}.\label{k2} 
\end{equation}  
Using above projection, eqn. (\ref{k1}) simplifies to: 
\begin{equation} 
\Gamma_{\hat + \hat - \hat 6 \hat 8}~\epsilon = \epsilon.\label{h1}
\end{equation}
The $\kappa$-symmetry projection of two intersecting $Dp$-branes at an 
arbitrary angle can be obtained by applying certain lorentz transformation 
$(\Lambda)$ as given in \cite{ortin}: 
 
\begin{equation}
\tilde\Gamma_a = \Gamma_b {\Lambda^b}_a = S^{-1} \Gamma_a S,
\end{equation}

where $\tilde\Gamma_a$ is the projection that depends on 
the detailed configuration of the second brane and $S$ is an element in 
$spin(1,9)$ that depends on $\Lambda$. In our case,  
$\tilde\Gamma_{(0)}$ is the projection for the  rotated $D3$-brane,
which can be expressed in terms of ${\Gamma_{(0)}}$,
the unrotated $D3$-brane projection: 

\begin{eqnarray}
\tilde\Gamma_{(0)} &=&  \Gamma_{\hat + \hat -}
( \cos\alpha~{\Gamma_{\hat 6}} +  \sin\alpha~{\Gamma_{\hat  5}}) 
( \cos\alpha~{\Gamma_{\hat 8}} +  \sin\alpha~{\Gamma_{\hat  7}}),\cr
& \cr
&=&  \Gamma_{\hat + \hat - \hat 6 \hat 8}~ 
e^{\alpha(\Gamma_{\hat 6 \hat 5} + \Gamma_{\hat 8 \hat 7})}.\label{k3}
\end{eqnarray}
Using eqn.(\ref{k1}) and (\ref{h1}) in eqn. (\ref{k3}) one gets 
the constraint : 
\begin{eqnarray}
[e^{\alpha~(\Gamma_{\hat 6 \hat 5} + \Gamma_{\hat 8 \hat 7})}- 1 ]
~\epsilon = 0
= \sin \alpha\Gamma_{\hat 6 \hat 5}
~e^{\alpha~\Gamma_{\hat 6 \hat 5}} (1 - \Gamma_
{\hat 5 \hat 6 \hat 7 \hat 8})~\epsilon.\label{k4}
\end{eqnarray}
For any arbitary angle $(\alpha)$ eqn.(\ref{k4}) can be satisfied only 
when
\begin{equation}
\Gamma_{\hat 5 \hat 6 \hat 7 \hat 8}~\epsilon = \epsilon = 
\Gamma_{\hat + \hat -\hat 5 \hat 7 }
\Gamma_{\hat + \hat -\hat 6 \hat 8}~\epsilon.
\end{equation}
Using eqn.(\ref{h1}) the above condition simplies to:
\begin{equation}
\Gamma_{\hat + \hat -\hat 5 \hat 7 }~\epsilon = \epsilon.\label{k5}
\end{equation}
But eqn.(\ref{k5}) is the supersymmetry condition 
associated with the rotated $D3$-brane along the $(x^+,x^-,x^5,x^7)$ 
directions. The supersymmetry conditions obtained in 
eqns. (\ref{h1}) and (\ref{k5}), commute with each other\cite{ortin},
which implies pull back of the $1/4$ 
supersymmetry to the worlvolume of the $D3$-branes. 
One can check that the probe analysis presented here
is also consistent with that of the flat space case \cite{douglas}.

Now we briefly outline the open string construction of the $D3$-brane 
system, presented from the supergravity point of view and from probe analysis, 
following \cite{douglas}. 
\noindent               
Boundary conditions are:
\begin{equation}
D3_1:\>\>\> \partial_{\sigma}x^{+,-,6,8} = 0, \>\>\>\>
\partial_{\tau}x^{i,5,7} = 0, \>\>\>\> i = {1....4}.
\label{B1}
\end{equation}
\begin{equation}
D3_2:\>\>\> \partial_{\sigma}(\cos \alpha~x^{6,8} {\mp} 
\sin\alpha~x^{5,7}) = 0,\>\>
\partial_{\tau}({\pm}\sin \alpha~x^{6,8} + \cos\alpha~x^{5,7}) = 0, 
\label{B2}
\end{equation}
with ${x^{+},x^{-}}$ and $x^i$ satisfying the usual Neumann and Dirichlet 
boundary conditions respectively.
In eqns. (\ref{B1}) and (\ref{B2}), $D3_1$ and $D3_2$ denotes 
the unrotated and the rotated branes.
Referring to the e.o.m. given in \cite{michi}, with
the above boundary conditions, it is straightforward to write down 
the mode expansion and hence the canonical commutation relation.
We, however skip the details here.  

We, therefore, have analyzed the existence and stability of a system 
of $D$-branes oriented by $SU(2)$ rotations in $NS-NS$ and $R-R$ pp-wave 
background, arising from $AdS_3\times S^3 \times R^4$, transverse to
the branes. We have also shown the presence of more $D$-brane 
bound states by applying $T$-duality symmetry. 
They preserve $1/16$ of the supersymmetries.
These configurations are also studied as the probe branes in 
the given background. They are shown to preserve $1/4$ supersymmetry 
of the world volume. It will be interesting 
to generalize these to arbitrary $SU(N)$ rotations. 
As there is an appropriate representation of world volume
gauge fields by means of the angle between the rotated 
branes, it may be interesting to study them further
in pp-wave background. By using these configurations one may like to
study the physics of black holes as well. 
\vskip .15cm
\noindent 
{\Large \bf Acknowledgement}
\vskip .15cm
\noindent
I thank Alok Kumar for suggesting the
problem. I would also like to thank Alok Kumar and
Kamal L. Panigrahi for many helpful discussions and the anonymous
referee for constructive suggestions.
\vskip .15cm
\noindent 
{\Large \bf Appendix A}
\def\theequation{A\arabic{equation}}
\setcounter{equation}{0}
\vskip .2cm
\noindent
Here we write down all the non vanishing Christoffels, 
derived from eqn. (\ref{d3-2}): 
\begin{eqnarray}
\Gamma^i_{+ +} &=& {({\mu x_i})^2 \partial_i X \over 4 (1 + X)^2} 
- {{\mu}^2 x_i \over (1 + X)}, ~~~~ \Gamma^i_{+ -} = 
{\partial_i X \over 4 (1 + X)^2}, \cr
& \cr
\Gamma^-_{- i} &=& \Gamma^+_{+ i} = 
 - {\partial_i X \over 4 (1 + X)}, ~~~~  \Gamma^-_{+ i}
= - ({\mu}^2 x_i), \cr
& \cr
\Gamma^j_{j i} &=& {\partial_i X \over 4 (1 + X)}, \cr
& \cr
\Gamma^5_{5 i} &=& {1 + X_1 \sin^2\alpha \over 
2\sqrt {1 + X}} \partial_i \bigg\{{1 + X_2 + X_1 \cos^2\alpha 
\over \sqrt {1 + X}}\bigg\} \cr
& \cr 
&-& { X_1 \cos\alpha \sin\alpha 
\over2 \sqrt {1 + X}} \partial_i \bigg\{{X_1 \cos\alpha \sin\alpha 
\over \sqrt {1 + X}}\bigg\} = \Gamma^7_{7 i}, \cr
& \cr
\Gamma^6_{6 i} &=& {1 + X_2 + X_1 \cos^2\alpha \over 
2\sqrt {1 + X}} \partial_i \bigg\{{1 + X_1 \sin^2\alpha 
\over \sqrt {1 + X}}\bigg\} \cr
& \cr 
&-& { X_1 \cos\alpha \sin\alpha 
\over2 \sqrt {1 + X}} \partial_i \bigg\{{X_1 \cos\alpha \sin\alpha 
\over \sqrt {1 + X}}\bigg\} = \Gamma^8_{8 i}, \cr
& \cr
\Gamma^5_{6 i} &=& - {1 + X_1 \sin^2\alpha \over 
2\sqrt {1 + X}} \partial_i \bigg\{{ X_1 \cos\alpha  \sin\alpha 
\over \sqrt {1 + X}}\bigg\} \cr
& \cr 
&+& { X_1 \cos\alpha \sin\alpha 
\over2 \sqrt {1 + X}} \partial_i \bigg\{{ 1 + X_1 \sin^2\alpha 
\over \sqrt {1 + X}}\bigg\} = - \Gamma^7_{8 i}, \cr
& \cr
\Gamma^6_{5 i} &=& - {1 + X_2 + X_1 \cos^2\alpha \over 
2\sqrt {1 + X}} \partial_i \bigg\{{ X_1 \cos\alpha  \sin\alpha 
\over \sqrt {1 + X}}\bigg\} \cr
& \cr 
&+& { X_1 \cos\alpha \sin\alpha 
\over2 \sqrt {1 + X}} \partial_i \bigg\{{ 1 + X_2 + X_1 \cos^2\alpha 
\over \sqrt {1 + X}}\bigg\} = - \Gamma^8_{7 i}.
\end{eqnarray}
Now we present some of the non vanishing   
spin connections. As the expressions are complicated and long, 
we mention only some of them:
\begin{eqnarray}
\omega_+^{\hat i\hat -} &=& { (\mu x_i)^2 \partial_i X \over 2(1 +
X)^{3/2}} - {{\mu}^2 x_i \over(1 + X)^{1/2}}, \cr
& \cr
\omega_+^{\hat + \hat i} &=& \omega_-^{\hat - \hat i} 
= { - \partial_i X \over 4(1 + X)^{3/2}}, \cr
& \cr
\omega_5^{\hat 5 \hat i} &=&  \bigg( 
{1 + X_1 \sin^2\alpha \over 
2\sqrt {1 + X}} \partial_i \bigg\{{1 + X_2 + X_1 \cos^2\alpha 
\over \sqrt {1 + X}}\bigg\} \cr
& \cr 
&-& { X_1 \cos\alpha \sin\alpha 
\over2 \sqrt {1 + X}} \partial_i \bigg\{{X_1 \cos\alpha \sin\alpha 
\over \sqrt {1 + X}}\bigg\}\bigg)
\bigg({1\over \sqrt{1 + X_1 \sin^2\alpha}}\bigg), \cr
& \cr
\omega_6^{\hat 6 \hat i} &=& \Bigg[\bigg({1 + X_2 + X_1 \cos^2\alpha \over 
2\sqrt {1 + X}} \partial_i \bigg\{{1 + X_1 \sin^2\alpha 
\over \sqrt {1 + X}}\bigg\} \cr
& \cr 
&-& { X_1 \cos\alpha \sin\alpha 
\over2 \sqrt {1 + X}} \partial_i \bigg\{{X_1 \cos\alpha \sin\alpha 
\over \sqrt {1 + X}}\bigg\}\bigg)(1 + X_1 \sin^2\alpha) \cr
& \cr
&+& \bigg( {1 + X_1 \sin^2\alpha \over 
2\sqrt {1 + X}} \partial_i \bigg\{{ X_1 \cos\alpha  \sin\alpha 
\over \sqrt {1 + X}}\bigg\} \cr
& \cr 
&-& { X_1 \cos\alpha \sin\alpha 
\over2 \sqrt {1 + X}} \partial_i \bigg\{{ 1 + X_1 \sin^2\alpha 
\over \sqrt {1 + X}}\bigg\} \bigg) X_1 \cos\alpha \sin\alpha \Bigg]\cr
& \cr
&\times& {1\over \sqrt {1 + X}  \sqrt {1 + X_1 \sin^2\alpha}}, \cr
& \cr
\omega_i^{\hat 5 \hat 6} &=& - {1 \over 
2} \partial_i \bigg\{{ X_1 \cos\alpha  \sin\alpha 
\over \sqrt {1 + X}}\bigg\} \cr
& \cr 
&+& { X_1 \cos\alpha \sin\alpha 
\over2(1 + X_1\sin^2\alpha)} \partial_i \bigg\{{ 1 + X_1 \sin^2\alpha 
\over \sqrt {1 + X}}\bigg\}, ~~ etc...
\end{eqnarray}
 


\end{document}